\documentclass[conference]{IEEEtran}
\IEEEoverridecommandlockouts
\usepackage{cite}
\usepackage{amsmath,amssymb,amsfonts}
\usepackage{algorithmic}
\usepackage{graphicx}
\usepackage{textcomp}
\usepackage{xcolor}
\usepackage{lipsum}
\def\BibTeX{{\rm B\kern-.05em{\sc i\kern-.025em b}\kern-.08em
    T\kern-.1667em\lower.7ex\hbox{E}\kern-.125emX}}
\begin{document}

\title{Web Security Investigation through Penetration Tests: A Case study of an Educational Institution Portal\\ 
}

\author{\IEEEauthorblockN{
D. Omeiza and J. Owusu-Tweneboah}
\IEEEauthorblockA{\textit{Carnegie Mellon University - Africa} \\
Kigali, Rwanda \\
\{domeiza, jowusu\}@africa.cmu.edu}
}

\maketitle

\begin{abstract}
Web security has become an important subject; many companies and organizations are becoming more security conscious as they build web applications to render online services and increase web presence. Unfortunately, many of these web applications are still susceptible to threats as they lack strong immunity to malicious attacks. This poses potential danger to the users of the sites and could also affect operations of the organizations or companies concerned. Educational institutions are not left out, their portals and websites hold vital information whose integrity is of utmost importance. Taking Carnegie Mellon University Africa's internship portal as case study, we carried out penetration tests to investigate web vulnerabilities and proffered possible remedies to the discovered vulnerabilities. Our result will inform educational institutions on better website security practices, especially in the African domain.
\end{abstract}


\section{Introduction}
About 24 million malware incidents targeting Africa were observed in 2016 \cite{b1}. As the internet community expands, web threats are on the rise and it is difficult to provide patches that aid in solving all security vulnerabilities. A vulnerability is a flaw or weakness in a system’s design, implementation or operation that could be exploited to violate the system’s security. A vulnerability is also a combination of three elements which involve the system flaw, attacker accessibility to the flaw and attacker efficiency to exploit the flaw \cite{b2}. Studies have shown that firewalls and anti-virus programs are not sufficient to provide effective system security \cite{b2}. There are attacks such as cross site scripting and many others which affect websites. Cross site scripting refers to the injection of malicious scripts into a legitimate website [3]. We investigate vulnerabilities and the safety level of students and companies’ data on the internship portal of Carnegie Mellon University Africa (CMUA).
\footnote {https://cmu-r.secure.force.com/Careers/InternshipHomePage}
The internship portal was purposely built to grant access to companies to upload internship vacancies available. Students can access the portal and are able to view company details, see vacancies available and get information on the application procedures. Students upload confidential information on the website. Thus, this site represents a modestly complex site with multiple types of data and different stakeholders. Understanding vulnerabilities will inform the CMUA Information Technology team, website administrators in other educational institutions, and anyone with knowledge on security policies and systems with the kind of security measures to put in place in their systems to avoid users’ data compromise.

In this work, penetration tests were carried out on the website to identify vulnerability of the website. However, attack recovery mechanisms were not investigated in this work.

Penetration testing examines the behavior of systems under extreme conditions to identify their weaknesses and vulnerabilities. Penetration testing can highlight issues in security using a variety of tools which are able to analyze the system, while others attack the system to exploit vulnerabilities. Penetration testing involves gathering information about the target system before the test (reconnaissance), identifying possible entry points, attempting to break in and reporting the findings \cite{b4}. Types of penetration tests include; external test, internal test, blind test and double-blind test. The choice depends on the environment where the test will be run. In this paper, we focus on external. External test aims to examine if illegitimate people can gain unauthorized access and what level of access can be gained while internal test simulates an attack from the inside behind the firewall by an authorized user with standard access privileges.

\section{Importance and Prior Work}
In related papers, authors mentioned how malicious attackers can edit customers’ information, how to perform tests to discover vulnerabilities and how to protect against attacks.
\subsection{Why secure web applications?}
\cite{b5} discussed web application security and the different techniques that can be used to establish potential vulnerabilities a web application might have. From the authors research, they found that the growth and evolution of the internet have impacted the way we generate and communicate many sensitive information.  The authors mentioned that e-commerce sites which produce data can be stolen or altered if they do not have the necessary safety measures for handling their use. The fact that computers worldwide are susceptible to attack by hackers or crackers able to compromise computer systems and steal valuable or delete a large part of the computer’s information is an issue which needs to be investigated. Some common attacks on web application mentioned by the authors include semantic URL; where the attacker modifies the URL to perform actions that are not originally planned to be handled properly by the server, cross-site scripting; which allows code injection by malicious web users, where false requests are sent using special tools and cross-site request forgery; which allows the attacker to send HTTP requests at will from the victim’s machine. This situation is essential to know if these systems and data networks are protected from any kind of intrusion. It is important that web applications are not only designed for the objectives for which they were created but they must also be designed to take care of the data and information generated in them.

\subsection{What kind of tests should be performed on websites?}
\cite{b4} presented different aspects of penetration testing including tools, attack methodologies and defense strategies. Different penetration tests were performed using a private network, devices and virtualized systems and tools. The authors performed attacks such as smartphone penetration testing, hacking phones Bluetooth, traffic sniffing, man-in-the-middle attack, spying and hacking remote PC via IP and open ports using advanced port scanner. The authors’ purpose was to explain the use of penetration testing and share concepts for understanding the test. Tools within the Kali Linux suite were used. The result showed a success rate of 14.29\% for a man-in-the-middle attack, and a remote PC hacking. The researchers also corroborated the fact that anti-viruses and anti-malware are not sufficient anymore to ensure security. They also expressed how important systems like weather systems could be easily violated through penetration testing tools. 

\cite{b6} presented an efficient integrated penetration testing tool to detect five of the top ten web application vulnerabilities to date. They also pointed out issues accompanied with some penetration testing tools. Issues such as generation of high rates of false positives/negatives, efficiency, performance and reliability of the results are concerns while using systems like Acunetix, ZAP and w3af for penetration testing. From their research, they gathered that lightweight penetration testing approaches which generate extremely low false positive/negative rates are required. Their Web Guardia (an integrated penetration testing system to detect web application vulnerabilities) is able to crawl through a given target web application and detect vulnerabilities such as SQLI, XSS, Unvalidated Redirects and Forwards, Insecure Direct Object References and Security Misconfigurations. Their tool uses a simple architecture which involves crawling, attacking analyzing and reporting modules. From their research, it was discovered that the use of Web Guardia can keep the generation of false positives and negatives at a minimum level because it is technically unfeasible to completely avoid generation of false positives and false negatives.

\subsection{How to mitigate Web Security Attacks?}
Protecting shared assets from unauthorized users has become more than necessity. \cite{b3} demonstrated that firewalls, antivirus software, Windows Defender and other prevention techniques will not suffice in preventing attacks. The authors examined different aspects of penetration testing to determine vulnerable applications and hosts using the Nmap and Metasploit frameworks. A virtualized system was used that included different versions of Windows and Linux OS. The results showed that tools such as Nmap and the Metasploit framework are effective tools for the discovery of open holes in a defense system or network. The test results indicated that vulnerable programs are the main source of the problem. Also, creating patches is not a good fix because same types of vulnerabilities are exploited daily through code injection, remote access, and denial of service.

\cite{b7} identified web defacement as one of the most common attacks on websites. The authors explained that there has been a gentle decline in the number of web defacement from 2013 to 2017. They proposed a Web Defacement and Intrusion Monitoring Tool (WDIMT) that could rapidly identify altered or deleted web pages. The proposed tool will also be used to regenerate the original content of a website after the website has been defaced, this is one of the strongest points of WDIMT. The tool uses four major commands to perform tasks, which are: Initialize, verify, force, and delete. Upon executing these commands, the WDIMT detects any defacement that may have occurred or is currently in progress. If defacement is noticed it re-uploads the original content of that web page. This tool may help to reduce the numbers and effects of web defacement attacks.

\section{Methodology}

\begin{figure}[htbp]
\centerline{\includegraphics[width=8.5cm,height=4cm]{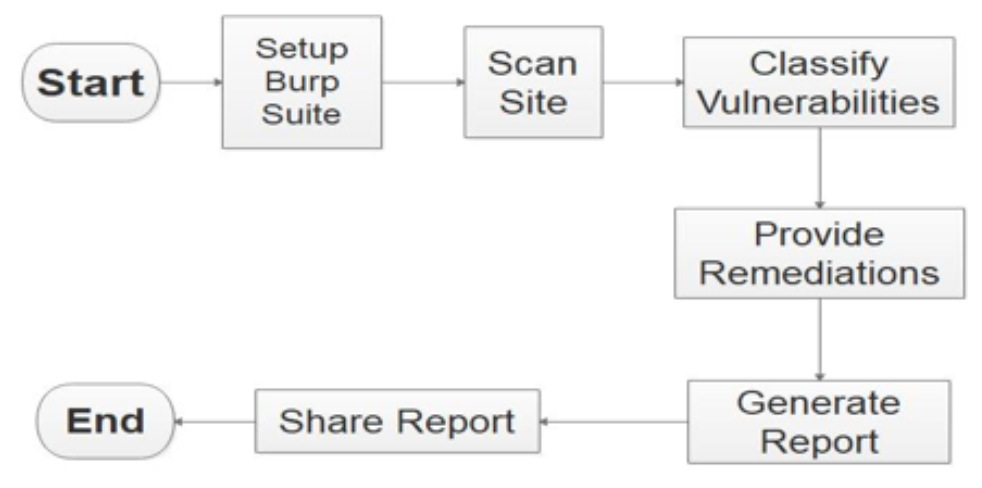}}
\caption{A high-level view of the step followed.}
\label{fig}
\end{figure}

Fig.1 above shows the conceptual view of the approach adopted. In getting a clue to how safe user data in Carnegie Mellon University Africa’s internship portal is, the answers to the following questions will be helpful. Is CMUA internship system secure? Are there vulnerabilities, and how severe are the vulnerabilities? How easy (in terms of time and resources committed) is it for an attacker to penetrate and compromise the system? The answers will be made clear in our results. The educational institutions as the targeted audience in this research will stand to benefit from the outcome of the research and may take actions where necessary to ensure maximum security. 

The work involved the use of BurpSuite for penetration testing of the selected website to investigate the degree of security of data in the website. Vulnerabilities discovered after the scan were classified by level of severity and certainty. In details we discuss the chosen tool, scope of the work, and approach.

\subsection{Tool Employed}
After analyzing the pros and the cons of existing penetration testing tools, BurpSuite was identified as the most suitable tool in the context of this research. BurpSuite is a cross-platform java application with several Web Application Security or penetration testing tools bundled into it to make a single suite.
Reasons for choosing BurpSuite is that it is easier to use, and basic tests do not require the writing of scripts. BurpSuite makes processes like intercept behavior configurations, application walk through, automated scanning and initial pilfering easier. Also, Kali Linux operating system comes with BurpSuite's free edition installed.

\subsection{Scope}
We limit this work to external penetration test operations which involve, scanning the website for SQL injection, cross-site scripting and other common vulnerabilities.
\subsection{Approach}
To investigate security through penetration testing on the internship portal, an approach that involved the use of BurpSuite for website scanning was used. The BurpSuite application can run on both Linux and Windows operating system. It provides functionalities for easy setup and automated scan.
Below are some of the features contained in BurpSuite as provided from BurpSuite’s documentation. Some of these tools will be employed in this research.
\begin{itemize} 
\item Spider: This can be used for automatically crawling an application, to discover its content and functionality.
\item Scanner: This is used to automatically scan HTTP requests to find security vulnerabilities.
\item Intruder: This allows you to perform customized automated attacks, to carry out all kinds of testing tasks.
\item Repeater: This is used to manually modify and reissue individual HTTP requests over and over.
\item Sequencer: This is used to analyze the quality of randomness in an application's session tokens.
\item Decoder: This lets you transform bits of application data using common encoding and decoding schemes.
\item Comparer: This is used to perform a visual comparison of bits of application data to find interesting differences.
The spider, scanner where explicitly used in our test operations, other features where employed implicitly during the run process.

Fig. 2 shows details of the steps (step 1 to step 7) we followed in using BurpSuite for the vulnerability analysis of the portal.
\end{itemize}

\begin{figure}[htbp]
\centerline{\includegraphics[width=8.3cm,height=4.2cm]{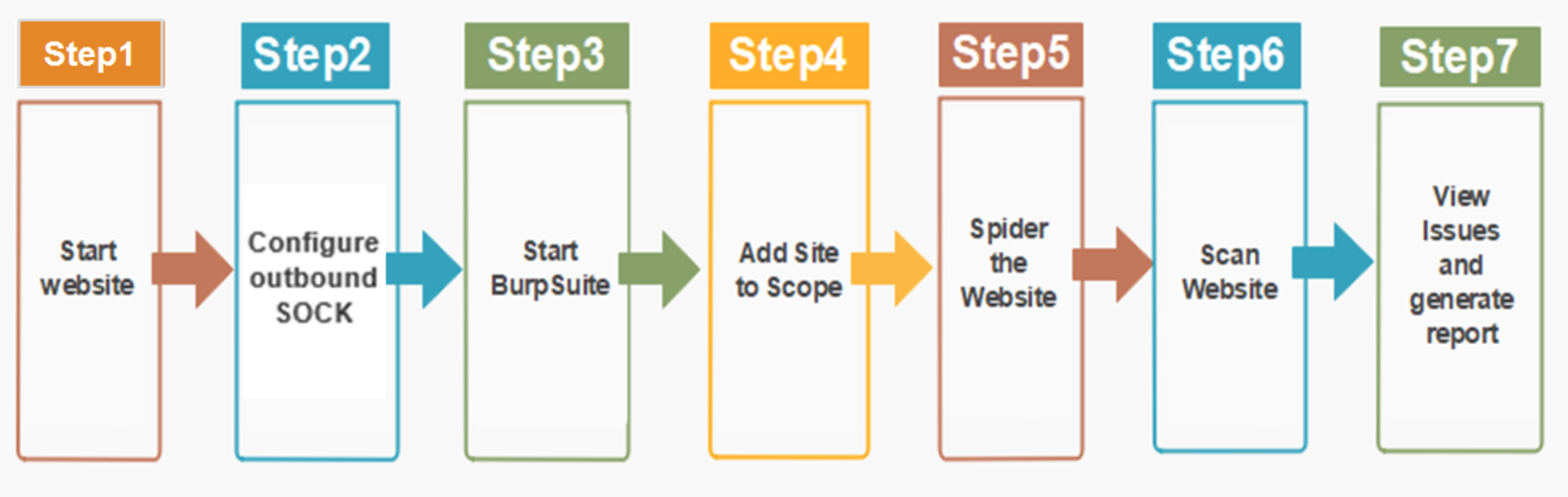}}
\caption{A high-level view of the step followed.}
\label{fig}
\end{figure}

\section{Penetration Testing Result and Discussion }
After scanning with BurpSuite, 16 certain low severity issues were discovered as shown in Table 1 below.

\begin{table}[]
    \centering
    \caption{Issues with severity and confidence level.}
    \begin{tabular}{l}
        \includegraphics[width=8.3cm,height=4cm]{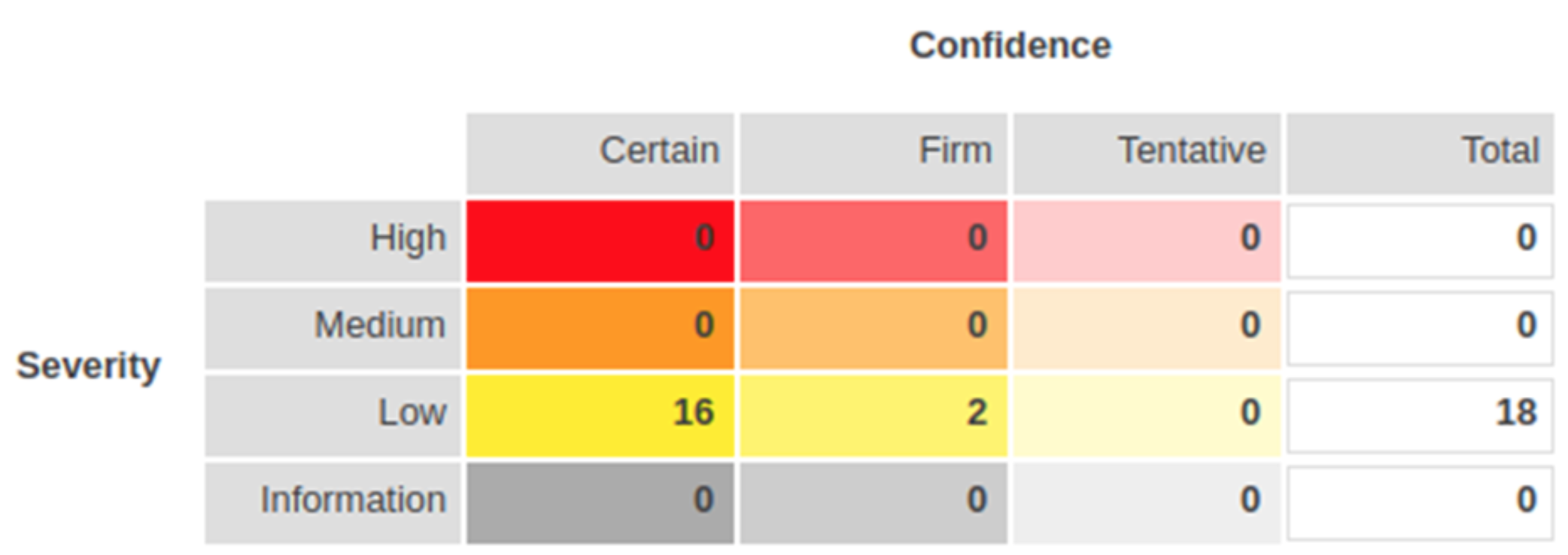}
    \end{tabular}
\end{table}
The confidence level indicates BurpSuite’s certainty of the issues discovered. The severity level indicates the impact a vulnerability has on the system if exploited. Vulnerabilities are hereby listed with corresponding remediation where necessary.
\subsection{Input returned in response (reflected)} 
There were two instances of this issue. It affected the Home and Contacts page of the website. This issue could lead to client-side vulnerabilities such as cross site scripting, open redirection, content spoofing and response header injection. In a case such as this, server-side vulnerabilities like SQL injection is easier. This vulnerability could be further classified as Improper Input Validation which leads to unintended inputs and Improper Encoding or Escaping of output which leads to the structure of the message not been preserved.

\textbf{Remediation:} Input should not be echoed at the application’s response.
\subsection {Cross-domain Referrer leakage}
There were three instances of this issue which occurred at the Contacts, Home and root page. The vulnerability was classified as Information Exposure vulnerability in the Common Vulnerabilities and Exposure (CVE) list.

\textbf{Remediation:} Applications should never transmit any sensitive information within the URL query string. In addition to being leaked in the Referrer header, such information may be logged in various locations and may be visible on-screen to untrusted parties.

\subsection {Cross-domain script include}
This occurs when an application includes scripts from an external domain. The script is executed by the browser within the security context of the invoking application and can function as the application’s own script and has access rights to the application data and perform actions within the context of the current user. The vulnerability was classified as Inclusion of functionality from untrusted control sphere vulnerability in the CVE list.

\textbf{Remediation:} Scripts should not be included from untrusted domains. Applications that consider the use of third-party scripts should copy the contents of the scripts to their own domain else re-implement the script’s functionality within the application code. 

\subsection{Password with autocomplete enabled}
There were two instances of this issue on the site. It occurred on the root page and the careers page. This function can be configured by the user and by applications that employ user credentials. This functionality allows the storage of user credentials on the local computer and is retrieved on future visits to the same application. The stored credentials can be captured by an attacker who gains control over the user’s computer. The vulnerability was classified as Inclusion of functionality from untrusted control sphere vulnerability in the CVE list.   

\textbf {Remediation:}
Include the attribute autocomplete="off" within the FORM tag (to protect all form fields) or within the relevant INPUT tags (to protect specific individual fields).
\subsection{Browser cross-site scripting filter disabled}
Some browsers, including Internet Explorer, contain built-in filters designed to protect against cross-site scripting (XSS) attacks. Applications can instruct browsers to disable this filter by setting the following response header: X-XSS-Protection: 0.
This behavior does not in itself constitute a vulnerability; in some cases, XSS filters may themselves be leveraged to perform attacks against application users. However, in typical situations, XSS filters do provide basic protection for application users against some XSS vulnerabilities in applications. The presence of this header should be reviewed to establish whether it affects the application's security posture.

\textbf{Remediation:} Review whether the application needs to disable XSS filters. In most cases you can gain the protection provided by XSS filters without the associated risks by using the following response header: X-XSS-Protection: 1; mode=block
\subsection{Client-side HTTP parameter pollution (reflected)} Client-side HTTP parameter pollution (HPP) vulnerabilities arise when an application embeds user input in URLs in an unsafe manner. An attacker can use this vulnerability to construct a URL that, if visited by another application user, will modify URLs within the response by inserting additional query string parameters and sometimes overriding existing ones. This may result in links and forms having unexpected side effects. For example, it may be possible to modify an invitation form using HPP so that the invitation is delivered to an unexpected recipient.

\textbf{Remediation:} Ensure that user input is URL-encoded before it is embedded in a URL.
\subsection{Email addresses disclosed} 
The presence of email addresses within application responses does not necessarily constitute a security vulnerability. Email addresses may appear intentionally within contact information, and many applications (such as web mail) include arbitrary third-party email addresses within their core content.

\textbf{Remediation:} Consider removing any email addresses that are unnecessary. Provide a form that generates the email server-side, protected by a CAPTCHA if necessary.
\subsection{Cacheable HTTPS response}

\textbf{Remediation:} Applications should return caching directives instructing browsers not to store local copies of any sensitive data. Often, this can be achieved by configuring the web server to prevent caching for relevant paths within the web root.

\section{Conclusion}
On investigating web security on educational institution web portal, Carnegie Mellon Africa’s internship portal was analyzed externally for vulnerabilities to determine the safety of user data on the site. The result from this analysis explains that the level of security of user data on the website is not sufficient. Although in context, the issues discovered are of low severity, the combination of such low severity issues become serious and demand attention. Remedies have been proffered for the vulnerabilities discovered. This stands to inform educational institutions on measures to assure secure websites and portals. Future work involves exploitation of XSS vulnerabilities by launching XSS attacks on a similar portal.


\begin{thebibliography}{00}
\bibitem{b1} V. Appiah, I. Kofi Nti, and O. Nyarko-Boateng, “Investigating websites and web application vulnerabilities: Webmaster’s Perspective,” Int. J. Appl. Inf. Syst., vol. 12, no. 3, pp. 10–15, 2017.

\bibitem{b2} S. P. Kadam, B. Mahajan, M. Patanwala, P. Sanas, and S. Vidyarthi, “Automated Wi-Fi penetration testing,” Int. Conf. Electr. Electron. Optim. Tech. ICEEOT 2016, pp. 1092–1096, 2016.

\bibitem{b3} 
O. Aslan and R. Samet, “Mitigating cyber security attacks by being aware of aulnerabilities and bugs,” 2017 Int. Conf. Cyberworlds, pp. 222–225, 2017. Suhl, Eds. New York: Academic, 1963, pp. 271-350.

\bibitem{b4} M. Denis, C. Zena, and T. Hayajneh, “Penetration testing: Concepts, attack methods, and defense strategies,” 2016 IEEE Long Isl. Syst. Appl. Technol. Conf. LISAT 2016, 2016.

\bibitem{b5} R. E. Lopez De Jimenez, “Pentesting on web applications using ethical hacking,” 2016 IEEE 36th Central American and Panama Convention (CONCAPAN XXXVI), no. 503, 2016.

\bibitem{b6} N. M. Vithanage and N. Jeyamohan, “WebGuardia - An integrated penetration testing system to detect web application vulnerabilities,” Proc. 2016 IEEE Int. Conf. Wirel. Commun. Signal Process. Networking, WiSPNET 2016, pp. 221–227, 2016.

\bibitem{b7} M. Masango, F. Mouton, P. Antony, and B. Mangoale, “Web Defacement and Intrusion Monitoring Tool : WDIMT,” 2017 International Conference on Cyberworlds (CW), 2017.
\end{thebibliography}
\end{document}